\documentclass[a4paper,11pt]{article}
\usepackage{graphicx}% Include figure files
\usepackage{dcolumn}% Align table columns on decimal point
\usepackage{bm}% bold math
\usepackage{color}

\begin{document}

\title{Search for $\Theta^{+}$ via $\pi^{-}p \to  K^{-}X$ reaction near production threshold}

\author{K. Miwa$^{a}$\footnote{Corresponding author, email: miwa9@nh.scphys.kyoto-u.ac.jp}, J.K. Ahn$^{f}$, K. Aoki$^{a}$, B.H. Choi$^{f}$, Y. Fukao$^{a}$, H. Funahashi$^{a}$,\\ T. Hayakawa$^{d}$, M. Hayata$^{a}$, T. Hibi$^{c}$, M. Ieiri$^{b}$, K. Imai$^{a}$, S.J, Kim$^{f}$,\\ K. Nakazawa$^{c}$,  H. Nishikawa$^{c}$, H. Okada$^{a}$,N. Saito$^{a}$, H.D. Sato$^{a}$, K. Shoji$^{a}$,\\ H. Takahashi$^{b}$, K. Taketani$^{a}$, K. Yamamoto$^{e}$, C. J. Yoon$^{a}$ 
}

%\date{\today}
\date{}
\maketitle
\begin{center}
{\it
$^{a}$Department of Physics, Kyoto University, Kyoto 606-8502, Japan\\
$^{b}$KEK, High Energy Accelerator Research Organization, Tsukuba 305-0801, Japan\\
$^{c}$Physics Department, Gifu University, Gifu 501-1193, Japan\\
$^{d}$Department of Physics, Osaka University, Toyonaka 560-0043, Japan\\
$^{e}$Department of Physics, Osaka City University, Osaka 558-8585, Japan\\
$^{f}$Department of Physics, Pusan National University, Pusan 609-735, Korea\\
}
\end{center}

\begin{abstract}
We have searched for $\Theta^{+}$ via $\pi^{-}p \to  K^{-}X$ reaction using
1.87 and 1.92 GeV/c $\pi^{-}$ beam at the K2 beam line of 
the KEK 12 GeV Proton Synchrotron.
In the missing mass spectrum at beam momentum of 1.92 GeV/$c$,
a bump has been found at 1530 MeV/$c^{2}$
which is consistent with the mass reported by several experiments.
The statistical significance of this bump, however, 
is only 2.5$-$2.7$\sigma$. 
Therefore we have derived the upper limit of 
$\Theta^{+}$ production cross section
via $\pi^{-}p \to  K^{-}\Theta^{+}$ reaction which is 
3.9$\mu$b at 90\% 
confidence level assuming that $\Theta^{+}$ is produced isotropically
in the center of mass system.
\end{abstract}

{\it PACS:} 12.39.Mk; 13.75.-n, 13.75.Gx; 14.20.-c \\
{\it Keywords:} Glueball and nonstandard multi-quark/gluon states, Hadron-induced low- and intermediate-energy reaction and scattering, Pion-baryon interactions, Baryons

\section{Introduction}
In the constituent quark model, mesons and baryons are described by
quark and anti-quark pair and three quarks respectively.
Many hadrons known so far are described by this model very well.
On the other hand, the existence of exotic baryons such as pentaquark and 
dibaryon has been suggested theoretically, because QCD requires only that
hadrons should be color singlet, 
but does not restrict the number of quarks \cite{Jaffe}.
Although many experimental efforts to search for these exotics were made,
there was no clear evidence.
%It might be one of the biggest mystery that these exotic particles 
%have not been observed experimentally.

It was about 30 years later from the establishment of QCD that
the first report about the exotic pentaquark, which is now called $\Theta^{+}$,
was made by SPring-8/LEPS collaboration \cite{LEPS}.
They showed a peak in mass distribution of $K^{+}n$ system via
$\gamma n \to K^{-} K^{+} n$ reaction where neutrons in $^{12}$C nuclei were
used as a target.
This discovery stimulated many physicists, and many works have been
done from both theoretical and experimental sides.
Diakonov et al., using the framework of chiral soliton model,
 regarded $\Theta^{+}$ as a member of anti-decuplet 
which is the third excited state of soliton field
and predicted that its mass was 1530 MeV/$c^{2}$ and the width was 15 MeV/$c^{2}$
by treating the known $N$(1710, 1/2$^{+}$) resonance as a member of 
this anti-decuplet \cite{Diakonov}. 
Jaffe and Wilczek proposed diquark-diquark-antiquark nature of $\Theta^{+}$
in the anti-decuplet plus octet representation of SU(3) \cite{JaffeWilczek}.
Many other theoretical studies such as constituent quark model,
QCD sum rules and lattice QCD were devoted to $\Theta^{+}$ \cite{Goeke}.

Experimentally, the observation at LEPS was immediately confirmed by several
experiments \cite{DIANA, CLAS, SAPHIR,CLAS2, nu, HERMES, ZEUS, COSY, SVD, C3H8}.
Recently, however, null results were reported from several high energy experiments 
where they searched for $\Theta^{+}$ with much higher statistics
\cite{HyperCP, HERA-B, ALEPH, BES, BABAR, CDF, SPHINX, PHENIX}.
Most recently, CLAS collaboration has reported some results
of a series of high statistical search for $\Theta^{+}$.
In \cite{CLAS3}, $\Theta^{+}$ was not observed and the upper limit of
production cross section via $\gamma p \to \bar{K}^{0} K^{+} n$ reaction 
was set to be 0.8 nb.
These negative results make the existence of $\Theta^{+}$ be in more
puzzling situation.
However Titov et al., using quark constituent coupling rules,
shows that the production of the $\Theta^{+}$ is suppressed relative to
the $\Lambda(1520)$ resonance by about three orders of magnitude for 
high energy experiments \cite{Titov}.
Therefore, in order to confirm the existence of $\Theta^{+}$, 
high statistics experiments at low energy region with hadronic reaction
become crucial.
Many experimental data of $\Theta^{+}$ are 
from photo-production experiments.
In general the $\Theta^{+}$ production cross section via hadronic reaction
is expected to be much larger than that via photo-induced reaction.
Now, physical properties such as spin, parity and width 
have not been determined experimentally yet.
In order to measure these values, we need more statistics.
From these viewpoints, the study of $\Theta^{+}$ production using 
meson beam such as $\pi^{-}$ and $K^{+}$ is essential.
Therefore we carried out an experiment to search for $\Theta^{+}$ via
$\pi^{-} p \to K^{-}\Theta^{+}$ reaction.
In this reaction, the threshold momentum is 1.71 GeV/$c$.
We used 1.87 and 1.92 GeV/$c$ $\pi^{-}$ beam.

The $\pi^{-}p \to K^{-} X$ reaction near the $\Theta^{+}$ production
threshold was studied in 1960s using a bubble chamber \cite{PIK}.
The main backgrounds of the ($\pi^{-}, K^{-}$) reaction are $\phi$ production,
$\Lambda(1520)$ production and 3-body phase space.
The cross sections of these reactions were measured to be
$30.0\pm 8.8 \mu {\rm b}, 20.8\pm 5.0 \mu {\rm b}$ and $\sim 25\mu$b respectively at
beam momentum from 1.8 to 2.2 GeV/$c$.
It is remarkable feature that in this momentum range 
the background is small because other channels do not open.
In this past experiment the invariant mass of $nK^{+}$ and $pK^{0}$ 
were surveyed. 
Any peak structure was not observed.
However numbers of $nK^{+}$ and $pK^{0}$ events were
only 86 and 249 respectively.

The understanding of production mechanism is quite important to understand
$\Theta^{+}$. 
Therefore a measurement of the production cross section with a simple reaction is
important experimentally.
Theoretically calculations with hadronic models using effective 
interaction Lagrangians and form factors were made 
by several authors \cite{Ko, Oh, Oh2, Hyodo}.
They try to understand the $\Theta^{+}$ production mechanism via
$\gamma N, NN, KN$ and $\pi N$ reactions near the production threshold 
comprehensively.
The theoretical calculations predict that the $\Theta^{+}$
production cross section of $\pi^{-} p \to  K^{-} \Theta^{+}$ reaction 
ranges from  several $\mu$b to several hundred $\mu$b.
However the parameters such as $g_{K^{*}N\Theta}$ used in
these calculations are not determined experimentally.
Therefore these calculation should be compared with experimental data.

In this letter we report the results of an experiment to search 
for $\Theta^{+}$ via $\pi^{-} p \to  K^{-} \Theta^{+}$ reaction.

\section{Experiment}

We have performed the E522 experiment at the K2 beam line of 
the KEK 12 GeV Proton Synchrotron in 2004. 
\hspace{-0.5ex}%
The main objective of this experiment was to search
for H-dibaryon resonance with ($K^{-}, K^{+}$) reaction. 
We searched for the enhancement at the threshold region of the double-$\Lambda$
system, which was first measured at KEK-PS E224 experiment \cite{Ahn}, with
much better statistics \cite{Yoon}.
Besides this reaction, we optionally took $\pi^{-} p \to  K^{-} X$ data,
because the $\Theta^{+}$ search
via mesonic reaction became crucial to confirm its existence
and the K2 beam line is a unique 
beam line which can provide up to 2 GeV/$c$ $\pi$ beam.

We used a $\pi^{-}$ beam of 1.87 and 1.92 GeV/$c$.
As a target, we used a scintillation fiber (SCIFI) target ($({\rm CH})_{n}$) 
and a bulk target of polyethylene ($({\rm CH}_{2})_{n}$).
The SCIFI target is 20 cm long, and is the same one used 
in the hyperon-nucleon 
scattering experiment (KEK-PS E289) \cite{Scifi}. 
It was mainly used to detect decay particles from 
$^{12}{\rm C}(K^{-}, K^{+} \Lambda \Lambda)$ reaction 
for the H-dibaryon resonance search.
For ($\pi^{-}, K^{-}$) data, we mainly used the 10 cm long polyethylene target
to enhance the  contribution from free protons.
The SCIFI target was also used to detect 
tracks of the charged particles other than $K^{-}$ produced by reactions.  
At the beam momentum of 1.87 GeV/$c$, 
2.9$\times 10^{9}$  and 3.0$\times 10^{9}$ $\pi^{-}$ beam particles 
were irradiated 
to the SCIFI and the polyethylene targets respectively.
At 1.92 GeV/$c$, 7.4$\times 10^{9}$ $\pi^{-}$  beam particles were irradiated
to only the polyethylene target.
For the calibration we took the following data.
In order to estimate
the contribution from carbon in the SCIFI and polyethylene targets, 
we took data with a carbon target. 
The ($\pi^{+}, K^{+}$) data were analyzed
to measure the $\Sigma^{+}$ peak position for the calibration of 
the missing mass spectrum.

\begin{figure}
\begin{center}
\includegraphics[width=12cm]{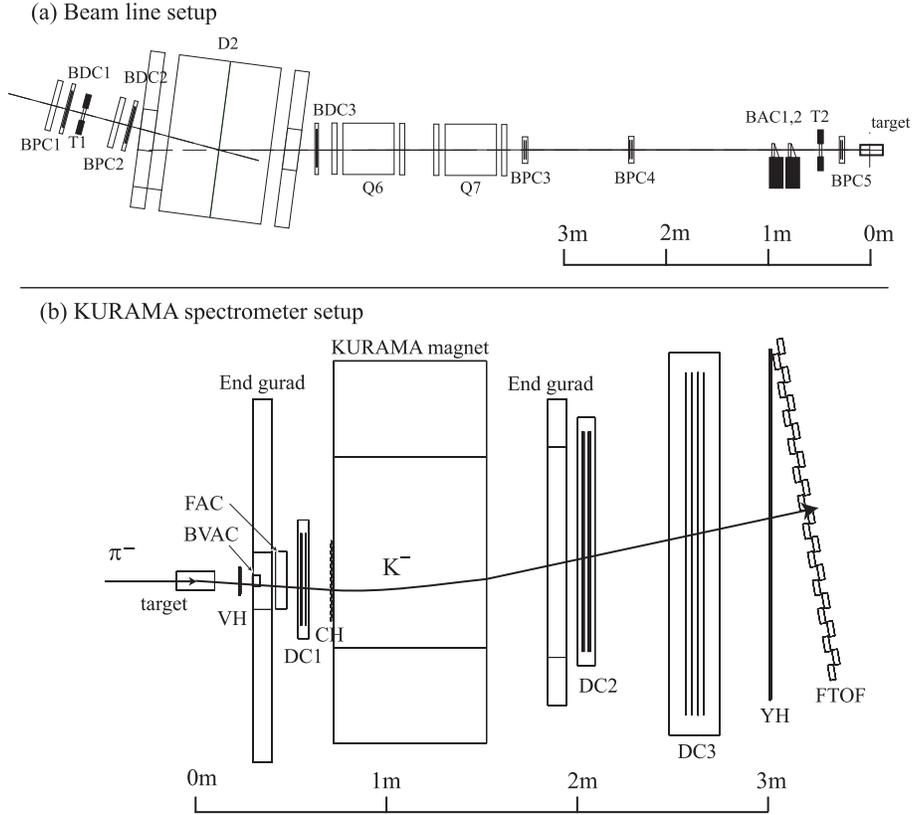}
\caption[]{Experimental setup. (a) shows the beam line spectrometer.
(b) shows the forward spectrometer.}
\label{KURAMAsp}
\end{center}
\end{figure}

The experimental setup consisted of two parts; one part was a beam line 
spectrometer to analyze momentum of each incident beam particle 
and the other part was
a forward spectrometer to detect scattered particles. 
The set up was almost similar to the one used 
at KEK-PS E373 experiment and 
detail of the experimental setup is described elsewhere \cite{E373}.

The K2 beam line is designed to transport charged particles up to 2.0 GeV/$c$.
Fig. \ref{KURAMAsp} (a) shows the beam line spectrometer.
The $\pi^{-}$ beam was bent by 15 degree at the bending magnet (D2) and
focused at the target by two quadrapole magnets (Q6, Q7).
The typical intensity of $\pi^{-}$ was 330k counts during 2 sec. spill
with a primary proton beam of $1.1 \times 10^{12}$.
Each beam particle was defined by the hit of T1 and T2 counters 
placed about 7.2m apart each other and 
$\pi^{-}$ was selected at a trigger by two aerogel 
cherenkov counters (BAC1,2), 
of which threshold velocity was 0.971, placed just upstream of the T2 counter.
The beam momentum was analyzed with 5 wire chambers placed upstream
and downstream of the D2 magnet. The momentum resolution is estimated
to be $\sigma(p) = 8.9$ MeV/$c$ from a simulation.
Between two quadrapole magnets
(Q6, Q7) and the target, 
3 proportional chambers were placed to measure the beam direction.
The timing for all detectors was determined by the T2 counter.

\begin{figure}[t]
\begin{center}
\includegraphics[width=7cm]{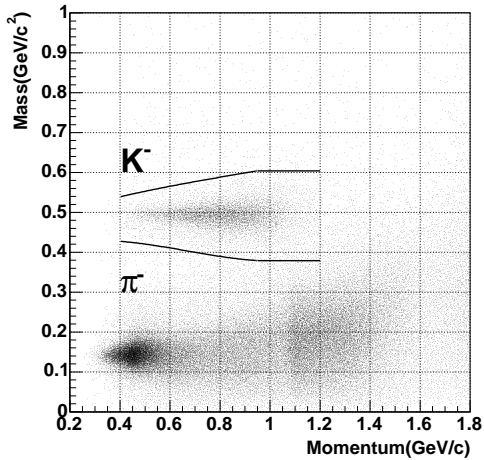}
\caption[]{A scattered plot of momentum and mass 
obtained from offline analysis.
By applying momentum dependent mass cut shown by 
solid lines,
$K^{-}$ mesons were selected cleanly.
}
\label{mass}
\end{center}
\end{figure}

The scattered particles were detected by the forward spectrometer
shown in Fig. \ref{KURAMAsp} (b).
The KURAMA magnet is 80 cm long and the magnetic field strength is 0.93T.
The momentum of each scattered particle was measured using
3 drift chambers (DC1,2,3) and scintillation hodoscopes (VH, CH) 
placed upstream and downstream of KURAMA.
The momentum resolution was 1.9 \% (r.m.s.) for 0.8 GeV/$c$ $K^{-}$.
The time-of-flight of each scattered particle was measured by a TOF wall
(FTOF) placed at end of the spectrometer.
A typical time resolution was 132 ps.
Two cherenkov counters (BVAC, FAC) were installed
between the target and KURAMA to veto $\pi^{-}$.
Besides this particle identification with these cherenkov counters
at the trigger level,
we selected the charge and momentum range of each scattered particle 
at 1st trigger level
using the hit combination of each segments of CH hodoscope and FTOF. 
Momentum of each particle was determined using the hit combination.
By combining time-of-flight information of the hit FTOF segment with this momentum information, 
the mass of each scattered particle was roughly calculated.
We selected $K^{-}$ with this mass trigger (MT) at 2nd trigger level.
This mass trigger rejected mainly $\pi^{-}$ 
which survived due to the inefficiency of
the cherenkov counters.
In the offline analysis, mass and momentum of each scattered particle
are calculated more accurately as shown in Fig. \ref{mass}.
The $K^{-}$ mesons are clearly identified.

\section{Analysis}

\begin{table}
\begin{center}
\caption{Summary of the analysis cuts and their efficiencies.
Errors are statistical ones.
This is a cut summary applied for the events 
where beam and outgoing particles are
found in analysis program.
The efficiencies of the track-finding routine are described in the next section
and listed in Table \ref{tab:summary_eff}. 
}
\label{tab:summary_ana_eff}
\ \\
%\begin{ruledtabular}
\begin{tabular}{cc}
\hline
\hline
cut  & efficiency(\%)\\
\hline
\multicolumn{2} {c}{$\pi^{-}$ selection}\\
\hline
Time-of-Flight cut & 93.8 $\pm$ 0.1 \\
beam $\chi^{2}$ cut & 84.6 $\pm$ 0.1 \\
beam momentum cut  & 94.8 $\pm$ 0.1 \\
\hline
\multicolumn{2} {c}{$K^{-}$ selection}\\
\hline
mass-momentum cut & $>99.7$ \\
Runge-Kutta $\chi^{2}$ cut & 72.7$\pm$0.2 \\
\hline
\multicolumn{2} {c}{vertex selection}\\
\hline
vertex cut & 92.8$\pm$ 0.5\\
closest distance cut & 97.8 $\pm$0.1 \\
\hline
total & 49.6$\pm$0.3 \\
\hline
\hline
\end{tabular}
%\end{ruledtabular}
\end{center}
\end{table}

We describe the analysis procedure to search for $\Theta^{+}$ from 
the missing mass spectrum of the $\pi^{-} p \to  K^{-} X$ reaction.
To obtain this spectrum, we applied the following cuts;
1) selection of $\pi^{-}$ in the beam and momentum analysis,
2) identification of the scattered $K^{-}$ and momentum analysis, and
3) selection of the reaction vertex point.
The efficiencies of these cuts in the analysis, 
which are described in detail in the following paragraphs,
are summarized 
in Table \ref{tab:summary_ana_eff}. 
The efficiencies of the track-finding routine are described in the next section
and listed in Table \ref{tab:summary_eff}. 

The incident $\pi^{-}$ was identified using
time-of-flight information between T1 and T2.
The contamination of $K^{-}$ was negligible, and we selected $\pm 3 \sigma$
region of its time resolution (70ps) as good events.
Each incident $\pi^{-}$ momentum was calculated by fitting the hit positions 
at the beam line chambers (BDC1,2,3 and BPC1,2)
with second order transfer matrix calculated by TRANSPORT \cite{transport}.
The $\chi^{2}$ distribution of this fitting was consistent with the expectation
except a tail towards large $\chi^{2}$. 
We selected C.L.=95\% region for $\chi^{2}$ cut whose efficiency was 84.6 \%
as the result of rejecting the large $\chi^{2}$ events.
The event whose beam momentum was less than 1.8 GeV/$c$ or larger than 
2.0 GeV/$c$ was rejected considering the momentum acceptance of the beam line.

Trajectories of outgoing particles were reconstructed by using the information
of hit positions at the drift chambers (DC1-3) and the hodoscopes (VH, CH) 
and of the field map in KURAMA.
At first, the local tracks upstream and downstream of KURAMA
were searched by the straight-line fittings, 
and the consistency between these local tracks were checked 
to reject the decay events of $K^{-}$.
Then we used Runge-Kutta method for the momentum analysis 
\cite{Runge-Kutta}.
The scattered $K^{-}$'s were selected by the momentum dependent mass cut
and we selected 3$\sigma$ region as shown in Fig \ref{mass}.
The contamination of $\pi^{-}$ after this cut was about 3\%.
In the ($\pi^{-}, K^{-}$) reaction, the momenta of scattered $K^{-}$ 
ranged from 0.4 to 1.1 GeV/$c$.
Therefore the effect of multiple scattering made the $\chi^{2}$ distribution of
Runge-Kutta tracking broader than the ideal one. 
To study the cut position of $\chi^{2}$ cut, we compared with 
the distribution obtained by the Monte Carlo simulation based on GEANT4 \cite{geant}
which included
all materials of the spectrometer and the electro-magnetic process, 
the hadoronic process and the decay process.
The simulated $\chi^{2}$ distribution reproduces real data except 
a long tail towards large $\chi^{2}$, not only for ($\pi^{-}, K^{-}$) 
reaction  but also
for ($\pi^{+}, K^{+}$) and ($K^{-}, K^{+}$) reactions where 
the typical momenta of $K^{+}$ were around 1.6 and 1.2 GeV/$c$ respectively.
From this study, we selected $\chi^{2} < 6.0$ region 
where the $\chi^{2}$ distribution was consistent with the simulation.

In the momentum reconstruction, energy loss effect in the materials 
of the spectrometer was not taken into account.
The energy loss in the target was  corrected using 
the reconstructed momenta of the incident and outgoing particles
with their directions and the calculated vertex positions. 

The vertex point was calculated by the closest distant point between
tracks of beam and outgoing particles.
We required the vertex point to be less than 80mm from the target center.
The beam size (1.4 $\times$ 1.3 cm$^{2}$) was small enough 
in comparison with the target size.
Because the SCIFI target enabled us to see particle trajectories as image data,
we could estimate the efficiency of this vertex cut precisely
by comparing the vertex position calculated by the spectrometer with 
one obtained in the image data.
The efficiency of this cut was estimated to be 92.8 $\pm$ 0.5 \%.
We also checked the distribution of the closest distance between
tracks of beam and outgoing particles at the vertex point.
The events where the closest distance was greater than 7mm,
 which corresponded to 3$\sigma$, were rejected
considering that
beam or outgoing particles reacted more than 2 times in the target.

\begin{figure}[t]
\begin{center}
\includegraphics[width=7cm]{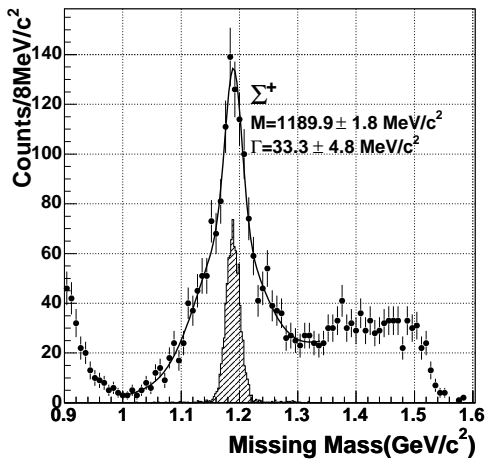}
\caption[]{Missing mass spectra of the ($\pi^{+}, K^{+}$) reaction with 
a polyethylene target. 
The beam momentum was adjusted so as to make the obtained $\Sigma^{+}$
peak consistent with the PDG value.
The hatched spectrum is the expected spectrum from the simulation.
The obtained peak width of 33.3 $\pm$ 4.8 MeV/$c^{2}$ (FWHM) is almost same with the expected value
28.3 MeV/$c^{2}$.
}
\label{missmass_pi+k}
\end{center}
\end{figure}

Fig. \ref{missmass_pi+k} shows the missing mass spectrum 
of the ($\pi^{+}, K^{+}$) reaction at the beam momentum of 1.92 GeV/$c$.
The peak due to $\Sigma^{+}$ is clearly observed.
The beam momentum was normalized so as to make the obtained $\Sigma^{+}$
peak consistent with the PDG value.
We fitted this spectrum with two Gaussian peaks assuming that 
the broad peak was attributed to quasi-free protons in carbon
and the narrow one was attributed to free protons.
The obtained width was 33.3$\pm$4.8 MeV/$c^{2}$ (FWHM) 
which was almost consistent with
the expected value of 28.3 MeV/$c^{2}$ from the simulation.
To estimate the missing mass resolution in the simulation, 
the position resolutions of 
the drift chambers, the momentum resolution of the beam spectrometer and
the effects of the energy loss and the multiple scattering in materials
for incident and outgoing particles were taken into account.
Using the same program code, the missing mass resolution for $\Theta^{+}$
was estimated to be 13.4 MeV/$c^{2}$ (FWHM). 
In $\Theta^{+}$ production, the momentum of the outgoing particle is much
lower than that in the $\Sigma^{+}$ production.
Therefore the missing mass resolution is better for $\Theta^{+}$.

\section{Results and Discussion}

\begin{figure}[]
\begin{center}
\includegraphics[width=8cm]{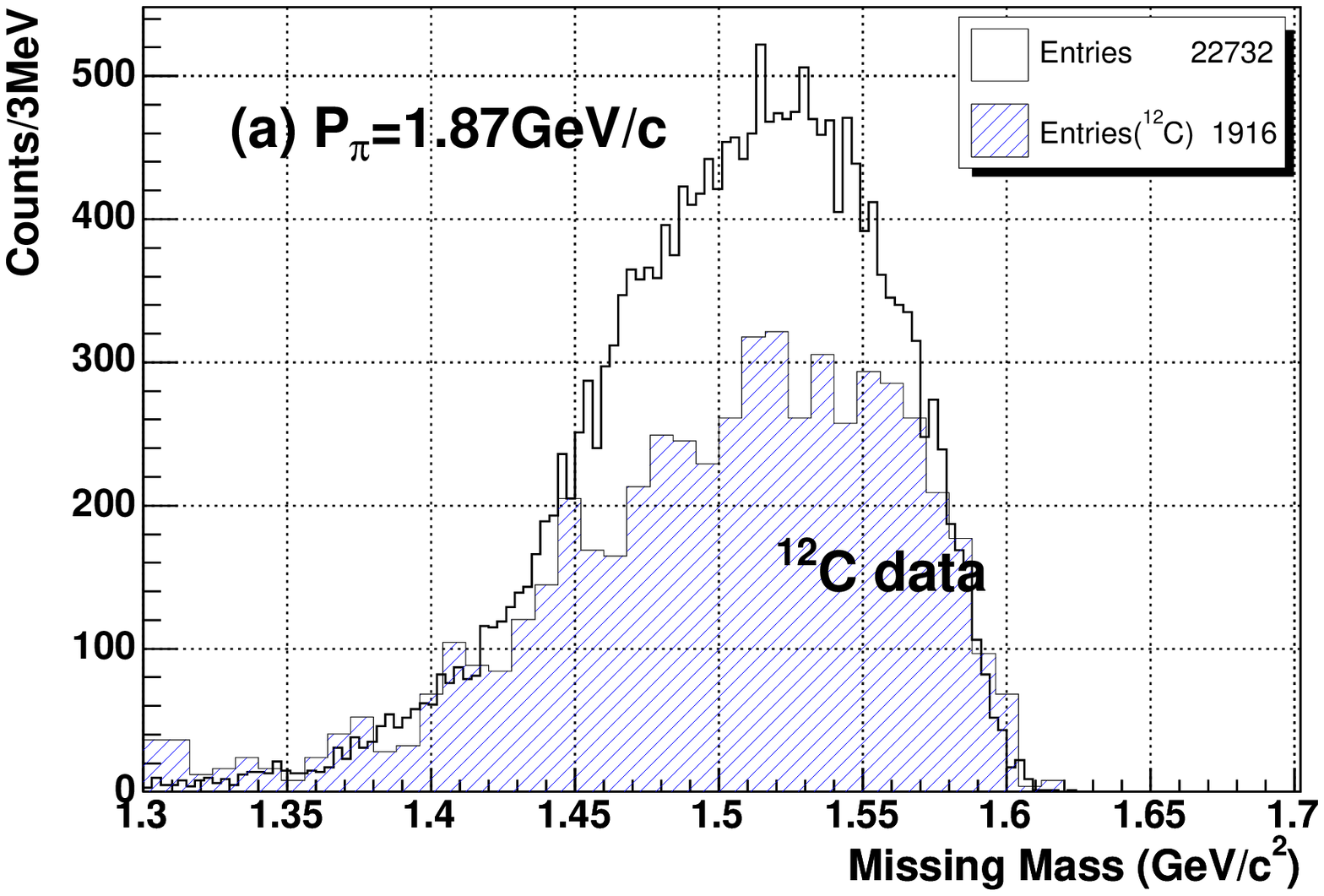}
\includegraphics[width=8cm]{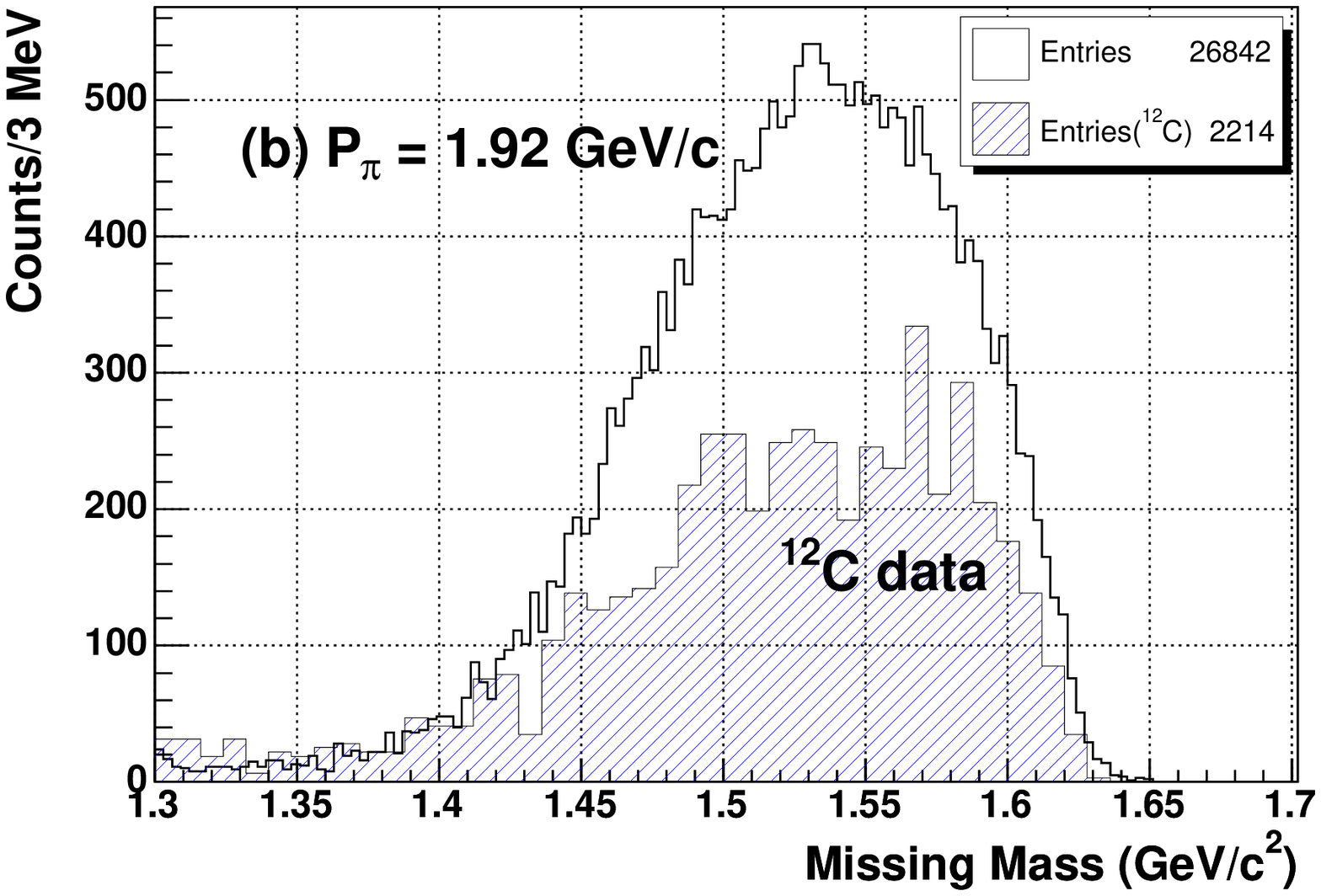}
\caption[]{Missing mass spectrum of the ($\pi^{-}, K^{-}$) reaction 
at 1.87 GeV/$c$ (a) and 1.92 GeV/$c$ (b).
The hatched histograms are the carbon target data
which are normalized by the number of target and beam particles,
and these histograms
represent the contribution from carbon nuclei in SCIFI and polyethylene targets.
The spectra where missing mass is less than 1.45 GeV/$c^{2}$ 
are well reproduced by these carbon target data, because
kinematically these events are dominantly from the carbon nuclei.
}
\label{missmass_pi-k}
%\end{figure}

%\begin{figure}
%\begin{center}
\includegraphics[width=9cm]{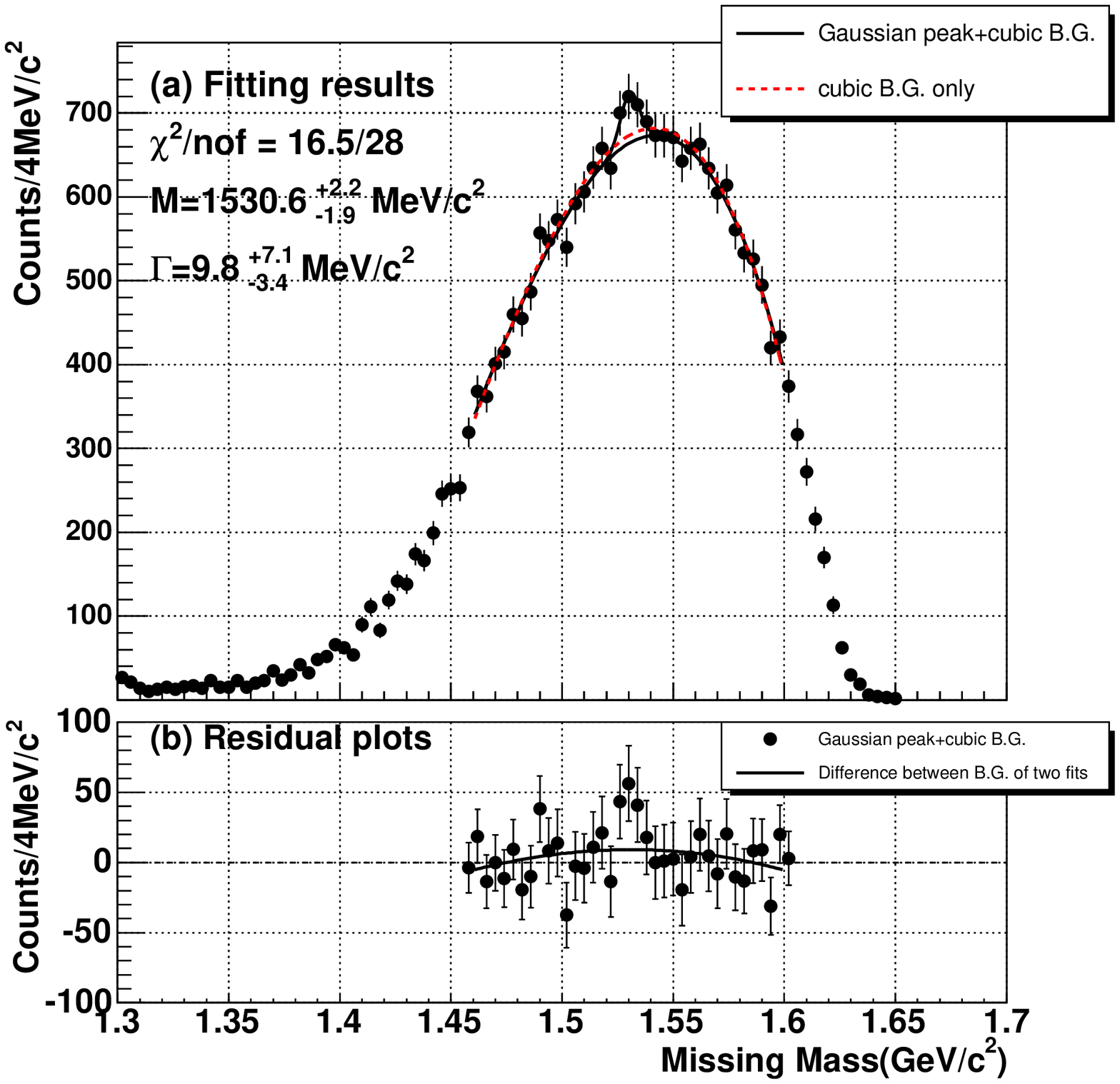}
\caption[]{Missing mass spectrum of the ($\pi^{-}, K^{-}$) reaction 
at 1.92 GeV/$c$.
(a) We fitted this spectrum with third order polynomial background 
and a Gaussian peak (solid line).
In this fitting, the width was a parameter 
and obtained to be $9.8^{+7.1}_{-3.4}$
MeV/$c^{2}$ (FWHM) which was consistent with the expected value of 
13.4 MeV/$c^2$ within the error.
We also executed the fitting with the fixed width of 13.4 MeV/$c^{2}$
(see text).
The dashed line represents the fitting result with only third order polynomial
background assuming that there is no peak structure.
(b) Residual plot from the background function obtained from the fitting
with third order polynomial background and a Gaussian peak.
The difference between backgrounds obtained from two fits, Gaussian peak
plus background and background only,
was represented by the solid line.
}
\label{fit_missmass_pi-k}
\end{center}
\end{figure}

The missing mass spectra for each beam momentum are shown in 
Fig. \ref{missmass_pi-k}.
In order to know the contribution from carbon nuclei 
in the SCIFI and polyethylene targets, 
carbon target data are also shown as the hatched spectra, 
which are normalized by the number of beam particle and 
the number of carbon nuclei in each target.
The statistics of the carbon target data is about ten times lower
than that of CH$_{2}$ data.
The net contribution from free protons is compatible with
the expectation from the cross sections of background reactions
measured in \cite{PIK}.
Because half of data at beam momentum of 1.87 GeV/$c$ was 
taken with the SCIFI target,
the contribution from free protons at 1.87 GeV/$c$ is smaller than 
that at 1.92 GeV/$c$.

Fig. \ref{missmass_pi-k} (a) shows the missing mass spectrum of
the $\pi^{-} p \to  K^{-} X$ reaction at 1.87 GeV/$c$.
In this spectrum, data for the SCIFI target and the polyethylene target
are combined.
Any peak was not observed in this spectrum.

Fig. \ref{missmass_pi-k} (b) shows the missing mass spectrum of
$\pi^{-} p \to K^{-} X$ reaction at 1.92 GeV/$c$.
It seems that there is a bump around 1.53 GeV/$c^2$.
However there is a possibility that the structure is only a statistical
fluctuation. 
We fitted this histogram with the background of a cubic function
and a Gaussian peak (see Fig. \ref{fit_missmass_pi-k}). 
The peak position is 
1530.6 $^{+2.2}_{-1.9}$(stat.)$^{+1.9}_{-1.3}$(syst.) MeV/$c^{2}$, 
the width is $\Gamma$=9.8$^{+7.1}_{-3.4}$ MeV/$c^{2}$ (FWHM)
and the count of the peak is 
139$^{+86}_{-67}$(stat.)$\pm$10(syst.).
The uncertainty of count resulting 
from varying the fitting range and binning 
is considered as the systematic error.
The obtained width is consistent with the expected 
value of 13.4 MeV/$c^{2}$ within the error.
We also fitted this spectrum with the fixed width of 13.4 MeV/$c^{2}$.
The count of this bump was obtained to be 
183$\pm$71(stat.)$\pm$10(syst.).
The statistical significance of this bump was considered by two expressions.
The first expression is the naive estimator 
$N_{s}^{2\sigma}/\sqrt{N_{s}^{2\sigma}+N_{b}^{2\sigma}}(\cong N_{s}^{2\sigma}/\sqrt{N_{b}^{2\sigma}})$ 
where $N_{s}^{2\sigma}$ is the peak count
within 2$\sigma$ region from the center and $N_{b}^{2\sigma}$ is the background
within the same region, 
and the significance is 2.7$\sigma$.
The second estimate of significance is given by $N_{s}/\sqrt{\Delta N_{s}}$,
where $N_{s}$ is full area of the bump from the fit and 
$\Delta N_{s}$ is its fully correlated uncertainty.
The significance of 2.5$\sigma$ is obtained by the second expression.

We fitted the histogram with only the background of a cubic function
assuming that there is no peak structure.
The fitting result is shown by a dashed line in Fig. \ref{fit_missmass_pi-k}.
The statistical significance of the bump from this background
is estimated to be 1.9$\sigma$  using the first expression,
$N_{s}^{2\sigma}/\sqrt{N_{s}^{2\sigma}+N_{b}^{2\sigma}}$.

The statistical significance 
obtained in the present experiment is not sufficient 
to claim this bump as the evidence of $\Theta^{+}$.
However it is quite important to estimate the upper limit of the production 
cross section of $\Theta^{+}$ via the $\pi^- p \to  K^- \Theta^{+}$ reaction. 
Therefore we have obtained the upper limit of the production cross section.
To derive the upper limit, the peak count obtained from the fitting with 
the fixed width is used at beam momentum of 1.92 GeV/$c$.
We used a single tail approach assuming that the peak count fluctuates based on Gaussian statistics.
Then the upper limit of the peak count is
$N_{s}+1.28\times \sqrt{\Delta N^{2}_{stat.}+\Delta N^{2}_{syst.}}$=274
at 90\% confidence level, where $N_{s}$ denotes the peak count obtained from 
the fitting and $\Delta N_{stat.}$, $\Delta N_{syst.}$ denote
the statistical and systematic errors respectively,
and we use this count for the following calculations.

In the missing mass spectrum at 1.87 GeV/$c$,
we could not find any obvious peak structure.
We estimated that the signal from $\Theta^{+}$($N_{\Theta^{+}}$) is less than 
1.28$\times \sqrt{N_{2\sigma}}$ at 90 \% confidence level, 
where $N_{2\sigma}$ 
represents the count in the missing mass spectrum corresponding to 
$\pm 2\sigma$ region from the peak position (1530.6 MeV/$c^{2}$) obtained
from 1.92 GeV/$c$ data.
We calculated that $N_{\Theta^{+}}$ were 62 and 52 for the SCIFI target and 
the polyethylene target, respectively.

The production cross section of $\Theta^{+}$ was calculated by 
the following equation.

%\begin{widetext}
\begin{eqnarray*}
  \left(
  \frac{d\sigma}{d\Omega}
  \right)
  &=& \frac{1}{N_{target}}\cdot \frac{1}{N_{beam}\cdot f_{beam} \cdot \epsilon_{K2}} \cdot \frac{1}{\epsilon_{DAQ}\cdot f_{K^{-}abs}} \cdot \frac{N_{\Theta^{+}}} {\epsilon_{track} \cdot f_{decay} \cdot f_{Cherenkov} \cdot \epsilon_{MT}     \cdot \epsilon_{ana} \cdot d\Omega} 
%  N_{target} &=& \frac{2 \cdot (\rho x) \cdot N_{Avo}}{14} ~({\rm for~ polyethylene~ target}), ~N_{target} = \frac{(\rho x) \cdot N_{Avo}}{13} ~({\rm for ~SCIFI ~target})
\end{eqnarray*}
%\end{widetext}

Here $N_{\Theta^{+}}, N_{beam}$ and $N_{target}$ represent the
number of $\Theta^{+}$, beam particles and protons in the  target.
The solid angle covered by KURAMA
spectrometer at laboratory frame is represented by d$\Omega$.
Others represent various efficiencies, and 
are summarized in Table\ref{tab:summary_eff}.

\begin{table}
\begin{center}
\caption{Summary of the various efficiencies for the calculation of the production cross section.}
\label{tab:summary_eff}
\ \\
%\begin{ruledtabular}
\begin{tabular}{c|c|c}
\hline
\hline
$f_{beam}$ & beam normalization factor  & 83.6 $\pm$ 1.3\% \\
$\epsilon_{K2}$ & track-finding efficiency of K2 beam line  & 72.7$\pm$ 2.0\% \\
$\epsilon_{track}$ & track-finding efficiency of scattered particle  & 84.6 $\pm$ 2.0 \% \\
$f_{Cherenkov}$ & Cherenkov overkilling factor  &  90.8$\pm$ 0.8\% \\
$f_{decay}$ & decay factor  & 57.6$\pm$ 0.1\% \\
$f_{K^{-}abs}$ & $K^{-}$ absorption factor  & 89.8$\pm$ 0.1\% \\
$\epsilon_{DAQ}$ & DAQ live time  & 93.5$\pm$ 0.2\% \\
$\epsilon_{MT}$ & efficiency of mass trigger  &  95.7$\pm$ 0.1\% \\
$\epsilon_{ana}$ & efficiency of analysis cut &  49.6$\pm$ 0.3\% \\
\hline
\hline
\end{tabular}
%\end{ruledtabular}
\end{center}
\end{table}

The coefficient $f_{beam}$ is the correction factor 
to obtain the number of real $\pi^{-}$ .
In this experiment, we could not distinguish $e^{-}$ and $\mu^{-}$ from $\pi^{-}$.
We referred the past experiment where a gas cherenkov counter was
used to distinguish $e^{-}$ and $\mu^{-}$ and estimated that
this contamination was 12.4\% \cite{Yamamoto, Kondo}.
We also calculated the reaction rate of $\pi^{-}$ in the target
with GEANT simulation and obtained to be 4\%.
Adding these value, we estimated that $f_{beam}$ was 83.6\%.

The efficiencies of track-finding routines used in the analysis program
for beam and scattered particles have to be estimated, because routines
have criteria to find tracks such as minimum number of hit chambers and
can not find out a part of tracks due to the inefficiency of the drift chambers
or multi charged hit events.
The coefficient $\epsilon_{K2}$ is the track-finding efficiency 
for beam particles
and obtained to be 72.7\%.
The reason for the inefficiency is mainly due to the dead channel of BDC3.
The coefficient $\epsilon_{track}$ is the track-finding efficiency 
for scattered particles and 
obtained to be 84.6\% by analyzing the data produced by the Monte Carlo 
simulation with
the same analysis program.
The validity of this estimation was checked using scattered proton events
taken with ($K^{-}, K^{+}$) trigger data.
The scattered protons could be selected by using information
of hit combination of CH and FTOF counters and of time-of-flight
without tracking.
We estimated the track-finding efficiency by analyzing
such pre-selected proton events.
Results obtained from the simulated events and pre-selected proton events
were consistent within 2.0\% which denoted the error of the efficiency.

The coefficient $f_{Cherenkov}$  represents the correction factor due to
the overkilling rate of the cherenkov counters (BVAC, FAC)
which was 9.2\%. Therefore $f_{Cherenkov}$ was obtained to be 90.8\%.
The coefficients, $f_{decay}$ and $f_{abs}$, represent 
the correction factors due to
the decay rate before arriving FTOF wall and the interaction rate of $K^{-}$ 
in the materials of the target and the forward spectrometer respectively.
These factors were also calculated with the Monte Carlo simulation 
based on GEANT4 and obtained to be 57.6\% and 89.8\% respectively.
The coefficients, $\epsilon_{DAQ}$ and $\epsilon_{MT}$, are the efficiencies of
the DAQ system and 2nd level mass trigger, 
and obtained to be 93.5\% and 95.7\% respectively.
The coefficient $\epsilon_{ana}$ is the analysis cut efficiency 
and summarized in Table\ref{tab:summary_ana_eff} and estimated to be 49.6\%.
Finally, the solid angle covered by the spectrometer ($d\Omega$) is 
0.141 $\pm$ 0.004 sr for $\pi^{-} p \to K^{-} \Theta^{+}$ reaction
assuming that the mass of $\Theta^{+}$ is 1530.6 MeV/$c^{2}$.
%The range of scattered angles at the labolatory system is $0 \deg < \Theta < 20\deg$ and the mean value of scattered angles is 8.2$\deg$.
Scattered angles at the laboratory system range from $0~\deg$ to $20~\deg$
and the mean value of scattered angles is 8.2 $\deg$.

\begin{figure}
\begin{center}
\includegraphics[width=8.5cm]{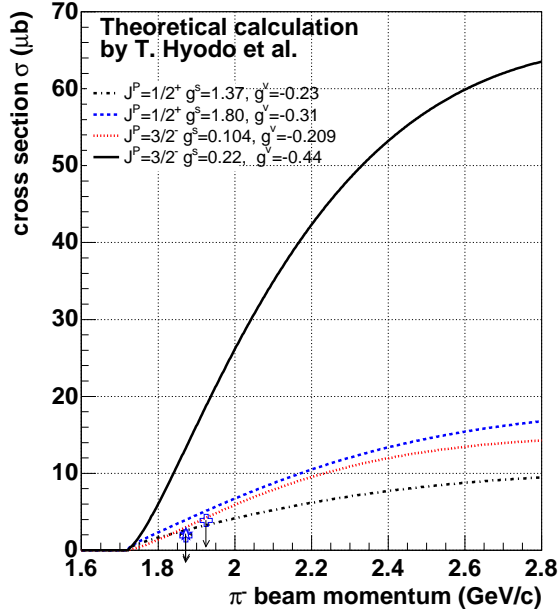}
\caption[]{The upper limits of the production cross section of 
$\pi^{-} p \to  K^{-} \Theta^{+}$ reaction at beam momentum of 1.87 and 1.92GeV/$c$.
The theoretical calculations by T. Hyodo et al. \cite{Hyodo2}
are also shown together.
They calculated the total cross section for $J^{P} = 1/2^{+}$
and $3/2^{-}$ using scalar and vector meson coupling constants, 
$g^{s}$ and $g^{v}$, respectively.
The dot-dashed and dashed lines are calculations for $J^{P}=1/2^{+}$
with ($g^{s}, g^{v}$) = (1.37, $-$0.23) and (1.80, $-$0.31) respectively.
The dotted and solid lines are calculations for $J^{P}=3/2^{-}$
with ($g^{s}, g^{v}$) = (0.104, $-$0.209) and (0.22, $-$0.44) respectively.
}
\label{cross_section}
\end{center}
\end{figure}

Using these values, we obtained the upper limit of the differential 
cross section via $\pi^{-} p \to K^{-} \Theta^{+}$ reaction 
for 1.92 GeV/$c$ data to be
$\frac{d\sigma}{d\Omega} = 2.9~ \mu {\rm b} / {\rm sr}$ 
at 90\% confidence level.
Assuming that $K^{-}$ is produced isotropically in the center of mass system,
10.4\% of $K^{-}$ is accepted by the spectrometer.
Therefore, if the $K^{-}$ is produced with s-wave, 
the upper limit of the total cross section 
is obtained to be 
${\sigma} = 3.9~ \mu {\rm b}$
 at 90\% confidence level.

We obtained the upper limit of the cross section from the 1.87 GeV/$c$ data 
as well as 1.92 GeV/$c$ data.
Because we used two different targets, we derived the upper limit 
for each target.
From the  SCIFI target data, the upper limit for $\frac{d\sigma}{d\Omega}$ 
and  $\sigma$ are obtained to be 1.7 $\mu$b/sr and 2.1 $\mu$b respectively.
From the polyethylene target data, the upper limit for $\frac{d\sigma}{d\Omega}$ 
and  $\sigma$ are obtained to be 1.6 $\mu$b/sr and 1.8 $\mu$b respectively.
These results from the two targets are consistent each other.

%These values are much smaller cross section than other hadron resonance and
%is also smaller than some theoretical calculations.
The theoretical calculations for this reaction have been done by
W. Liu and C. M. Ko \cite{Ko} and Y. Oh et al. \cite{Oh}.
These theoretical calculations depend on the values of the form factor 
and coupling constants.
In Ref. \cite{Ko}, Liu and Ko calculated the cross section 
taking into account only the s-channel diagrams.
They used $g_{KN \Theta}=4.4$, which corresponds to 20 MeV/$c^{2}$ width of $\Theta^{+}$, and a cutoff value of $\Lambda=0.5$ GeV for the form factor.
They predict that the cross section is about 50 $\mu$b.
Y. Oh et al. calculated the cross section taking into account
the s-channel diagrams and t-channel diagrams where
the $K^{*}$ is exchanged.
They used $g_{KN \Theta}$=2.2, which corresponds to 5 MeV/$c^{2}$ width of $\Theta^{+}$, and the same cutoff value used by C. M. Ko et al..
Because there is no information about $g_{K^{*}N \Theta}$,
they used several values from $-$2.2 to 2.2 as $g_{K^{*}N \Theta}$.
The calculated cross section ranges from about 2 $\mu$b to 190 $\mu$b.
Present results are quite smaller than the theoretical calculations and
gives strong constraint to these unknown parameters.
Recently, a theoretical study of production mechanism via hadronic reactions
has been done vigorously by T. Hyodo et al. \cite{Hyodo2}.
Fig. \ref{cross_section} shows the obtained upper limit of the cross section 
for each $\pi^{-}$ beam momentum together with their theoretical calculations.
They took particular note of the importance of two meson coupling, and 
calculated the total cross sections of the reaction 
$\pi^{-} p \to K^{-} \Theta^{+}$ and $K^{+} p \to \pi^{+} \Theta^{+}$
in case of $J^{P} = 1/2^{+}$ and $3/2^{-}$.
They obtained the scalar and vector coupling constants of $\Theta K \pi N$ using the SU(3)
symmetry from the decay width of $N^{*}(1710) \to \pi\pi N$.
The obtained coupling constants have uncertainty due to
the experimental uncertainties of the branching ratio.
They restricted  the coupling constants to be consistent with our present 
results for 
the dotted, dot-dashed and dashed lines in Fig. \ref{cross_section}.
The ratio of cross sections of $\pi^{-}$ and $K^{+}$ induced reactions 
can be calculated more reliably than the absolute value of the cross section.
By using the coupling constant which explain the present data,
they find that the ratio is very different for two $J^{P}$ assignments.
In the case of $J^{P} = 1/2^{+}$ the ratio of the cross section,
$\sigma(K^{+}p \to \pi^{+}\Theta^{+})/\sigma(\pi^{-}p \to K^{-}\Theta^{+})$, 
is $\sim$ 50,
while in the case of $J^{P} = 3/2^{-}$ it is $\sim$ 3.3.
An experiment to search for $\Theta^{+}$ 
via  $K^{+} p \to \pi^{+} \Theta^{+}$ has being performed
at KEK (KEK-PS E559).
This experiment together with the present results provides
deeper understanding on the existence of $\Theta^{+}$.
%\section{summary}

We have searched for $\Theta^{+}$ via $\pi^{-} p \to K^{-} X$ reaction
with $\pi^{-}$ beams of 1.87 and 1.92 GeV/$c$.
In the missing mass spectrum at the beam momentum of 1.87 GeV/$c$, 
no clear peak was found.
At 1.92 GeV/$c$,
a bump has been found at 1530.6 MeV/$c^{2}$.
The statistical significance of this bump is 2.5$-$2.7$\sigma$
which is not sufficient to claim this bump as the evidence of $\Theta^{+}$ .
We have obtained the upper limit of 
$\Theta^{+}$ production cross section
via $\pi^{-}p \to  K^{-}\Theta^{+}$ reaction 
at 90\% 
confidence level assuming that $\Theta^{+}$ is produced isotropically
in the center of mass system.
The upper limit have been obtained to be 1.8 and 3.9 $\mu$b at beam momenta
of 1.87 and 1.92 GeV/$c$, respectively.

\section{Acknowledgments}
We would like to express our thanks to staffs of KEK PS
and beam channel group for their
support to provide beam with the excellent condition during the experiment.
We also acknowledge to T. Hyodo for the theoretical discussion.
One of the authors (K. M.) thanks to the Japan Society for 
the Promotion of Science (JSPS) for support.
The work of J.K.Ahn was supported by a Korea Research Foundation
grant (KRF-2003-015-C00130).
This work was supported by the Grant-in-Aid for Specially Promoted Research 
(No.15001001) from the
Ministry of Education, Culture, Science and Technology, Japan.

\end{document}